\documentclass[aps,preprint, floatfix]{revtex4}

\usepackage{graphicx,color}
\usepackage{psfrag}
\usepackage{amssymb}
\usepackage{eso-pic}
\usepackage[dvips,letterpaper]{geometry} 
\usepackage{amsmath}

\begin{document}

%		DEFINITIONS FOR TEX
%
%%%%%%%%%%%%%%%%%%%%%%%%%%%%%%%%%%%%%%%%%%%%%%%%%%%%%%%%%%%%%%%
%
%
%\def\e{\'e}
%\def\ee{\`e}
%%%%%%%%%%%%%%%%%%%DEFINITIONS%%%%%%%%%%%%%%%%%%%%%%%%%%%%%%%%%
%
\def\oti{{\otimes}}
\def\lb{ \left[ }
\def\rb{ \right]  }
\def\tilde{\widetilde}
\def\bar{\overline}
\def\hat{\widehat}
\def\*{\star}
\def\[{\left[}
\def\]{\right]}
\def\({\left(}		\def\BL{\Bigr(}
\def\){\right)}		\def\BR{\Bigr)}
	\def\BBL{\lb}
	\def\BBR{\rb}
%
%%%%%%%%%%%%%%%%%%%%%%%%%%%%%%%%%%%%%%%%%%%%%%%%%%%%%%%%%%%%%%%
%
\def\zb{{\bar{z} }}
\def\zbar{{\bar{z} }}
\def\frac#1#2{{#1 \over #2}}
\def\inv#1{{1 \over #1}}
\def\half{{1 \over 2}}
\def\d{\partial}
\def\der#1{{\partial \over \partial #1}}
\def\dd#1#2{{\partial #1 \over \partial #2}}
\def\vev#1{\langle #1 \rangle}
\def\ket#1{ | #1 \rangle}
\def\rvac{\hbox{$\vert 0\rangle$}}
\def\lvac{\hbox{$\langle 0 \vert $}}
\def\2pi{\hbox{$2\pi i$}}
\def\e#1{{\rm e}^{^{\textstyle #1}}}
\def\grad#1{\,\nabla\!_{{#1}}\,}
\def\dsl{\raise.15ex\hbox{/}\kern-.57em\partial}
\def\Dsl{\,\raise.15ex\hbox{/}\mkern-.13.5mu D}
%
%%%%%%%%%%%%%%%%%%%%GREEK LETTERS%%%%%%%%%%%%%%%%%%%%%%%%%%%%%%
%
%\def\th{\theta}		\def\Th{\Theta}
\def\ga{\gamma}		\def\Ga{\Gamma}
\def\be{\beta}
\def\al{\alpha}
\def\ep{\epsilon}
\def\vep{\varepsilon}
\def\la{\lambda}	\def\La{\Lambda}
\def\de{\delta}		\def\De{\Delta}
\def\om{\omega}		\def\Om{\Omega}
\def\sig{\sigma}	\def\Sig{\Sigma}
\def\vphi{\varphi}

%
%%%%%%%%%%%%%%%%%%%CALIGRAPHIC LETTERS%%%%%%%%%%%%%%%%%%%%%%%%%
%
\def\CA{{\cal A}}	\def\CB{{\cal B}}	\def\CC{{\cal C}}
\def\CD{{\cal D}}	\def\CE{{\cal E}}	\def\CF{{\cal F}}
\def\CG{{\cal G}}	\def\CH{{\cal H}}	\def\CI{{\cal J}}
\def\CJ{{\cal J}}	\def\CK{{\cal K}}	\def\CL{{\cal L}}
\def\CM{{\cal M}}	\def\CN{{\cal N}}	\def\CO{{\cal O}}
\def\CP{{\cal P}}	\def\CQ{{\cal Q}}	\def\CR{{\cal R}}
\def\CS{{\cal S}}	\def\CT{{\cal T}}	\def\CU{{\cal U}}
\def\CV{{\cal V}}	\def\CW{{\cal W}}	\def\CX{{\cal X}}
\def\CY{{\cal Y}}	\def\CZ{{\cal Z}}

\def\rvac{\hbox{$\vert 0\rangle$}}
\def\lvac{\hbox{$\langle 0 \vert $}}
\def\comm#1#2{ \BBL\ #1\ ,\ #2 \BBR }
\def\2pi{\hbox{$2\pi i$}}
\def\e#1{{\rm e}^{^{\textstyle #1}}}
\def\grad#1{\,\nabla\!_{{#1}}\,}
\def\dsl{\raise.15ex\hbox{/}\kern-.57em\partial}
\def\Dsl{\,\raise.15ex\hbox{/}\mkern-.13.5mu D}
%
%%%%%%%%%%%%%%%%%%%%GREEK LETTERS%%%%%%%%%%%%%%%%%%%%%%%%%%%%%%
%
%%%%%%%%%%%%%%% MATH CHARACTERS %%%%%%%%%%%%%%%%%%%%%%%%%%%%
%
\font\numbers=cmss12
%\font\numbers=cmu10 scaled\magstep1
\font\upright=cmu10 scaled\magstep1
\def\stroke{\vrule height8pt width0.4pt depth-0.1pt}
\def\topfleck{\vrule height8pt width0.5pt depth-5.9pt}
\def\botfleck{\vrule height2pt width0.5pt depth0.1pt}
\def\Zmath{\vcenter{\hbox{\numbers\rlap{\rlap{Z}\kern
0.8pt\topfleck}\kern 2.2pt
                   \rlap Z\kern 6pt\botfleck\kern 1pt}}}
\def\Qmath{\vcenter{\hbox{\upright\rlap{\rlap{Q}\kern
                   3.8pt\stroke}\phantom{Q}}}}
\def\Nmath{\vcenter{\hbox{\upright\rlap{I}\kern 1.7pt N}}}
\def\Cmath{\vcenter{\hbox{\upright\rlap{\rlap{C}\kern
                   3.8pt\stroke}\phantom{C}}}}
\def\Rmath{\vcenter{\hbox{\upright\rlap{I}\kern 1.7pt R}}}
\def\Z{\ifmmode\Zmath\else$\Zmath$\fi}
\def\Q{\ifmmode\Qmath\else$\Qmath$\fi}
\def\N{\ifmmode\Nmath\else$\Nmath$\fi}
\def\C{\ifmmode\Cmath\else$\Cmath$\fi}
\def\R{\ifmmode\Rmath\else$\Rmath$\fi}
%%%%%%%%%%%%%%%%%%%%%%%%%%%%%%%%%%%%%%%%%%%%%%%%%%%%%%%%%%%%%%%%%
 %%%%%%%%%%%%%%%%%% END OF DEFINITIONS %%%%%%%%%%%%%%%%%%%%%%
 %%%%%%%%%%%%%%%%%%%%%%%%%%%%%%%%%%%%%%%%%%%%%%%%%

\def\barray{\begin{eqnarray}}
\def\earray{\end{eqnarray}}
\def\beq{\begin{equation}}
\def\eeq{\end{equation}}

\def\n{\noindent}

\def\Tr{\rm Tr} 
\def\xvec{{\bf x}}
\def\kvec{{\bf k}}
\def\kvecp{{\bf k'}}
\def\omk{\om{\kvec}} 
\def\dk#1{\frac{d\kvec_{#1}}{(2\pi)^d}}
\def\2pid{(2\pi)^d}
\def\ket#1{|#1 \rangle}
\def\bra#1{\langle #1 |}
\def\vol{V}
\def\adag{a^\dagger}
\def\rme{{\rm e}}
\def\Im{{\rm Im}}
\def\pvec{{\bf p}}
\def\fermiS{\CS_F}
\def\cdag{c^\dagger}
\def\adag{a^\dagger}
\def\bdag{b^\dagger}
\def\vvec{{\bf v}}
\def\muhat{{\hat{\mu}}}
\def\vac{|0\rangle}
\def\pcut{{\Lambda_c}}
\def\chidot{\dot{\chi}}
\def\gradvec{\vec{\nabla}}
\def\psitilde{\tilde{\Psi}}
\def\psibar{\bar{\psi}}
\def\psidag{\psi^\dagger} 
\def\m{m_*}
\def\up{\uparrow}
\def\down{\downarrow}
\def\Qo{Q^{0}}
\def\vbar{\bar{v}}
\def\ubar{\bar{u}}
\def\smallhalf{{\textstyle \inv{2}}}
\def\smallsqrt{{\textstyle \inv{\sqrt{2}}}}
\def\rvec{{\bf r}}
\def\avec{{\bf a}}
\def\pivec{{\vec{\pi}}}
\def\svec{\vec{s}} 
\def\phivec{\vec{\phi}}
\def\daggerc{{\dagger_c}}
\def\Gfour{G^{(4)}}
\def\dim#1{\lbrack\!\lbrack #1 \rbrack\! \rbrack }
\def\qhat{{\hat{q}}}
\def\ghat{{\hat{g}}}
\def\nvec{{\vec{n}}}
\def\bull{$\bullet$}
\def\ghato{{\hat{g}_0}}
\def\r{r}
\def\deltaq{\delta_q}
\def\gcharge{g_q}
\def\gspin{g_s}
\def\deltas{\delta_s}
\def\gQC{g_{AF}} 
\def\ghatqc{\ghat_{AF}}
\def\xqc{x_{AF}}
\def\mhat{\hat{m}}
\def\xup{x_2}
\def\xdown{x_1}
\def\sigmavec{\vec{\sigma}}
\def\xopt{x_{\rm opt}}
\def\Lambdac{{\Lambda_c}}
\def\angstrom{{{\scriptstyle \circ} \atop A}     }
\def\AA{\leavevmode\setbox0=\hbox{h}\dimen0=\ht0 \advance\dimen0 by-1ex\rlap{
\raise.67\dimen0\hbox{\char'27}}A}
\def\ratio{\gamma}
\def\Phivec{{\vec{\Phi}}}
\def\singlet{\chi^- \chi^+} 
\def\mhat{{\hat{m}}}

\def\Im{{\rm Im}}
\def\Re{{\rm Re}}

\def\xstar{x_*}

\def\sech{{\rm sech}}

\def\Li{{\rm Li}}

\def\dim#1{{\rm dim}[#1]}

\def\ep{\epsilon}

\def\free{\CF}

\def\Fhat{\digamma}

\def\ftilde{\tilde{f}}

\def\muphys{\mu_{\rm phys}}

\def\xitilde{\tilde{\xi}}

\def\CI{\mathcal{I}}

\def\nhat{\hat{n}}

\def\ef{\epsilon_F}

\def\as{a_s}

\def\diffk{|\kvec - \kvec' |}

\def\dk#1{\frac{d^3 #1}{(2\pi)^3}}

\title{Quantum  Bose and Fermi gases  with large negative scattering length in the 2-body S-matrix approximation}
\author{Andr\'e  LeClair$^{1,2}$,  Edgar Marcelino$^1$,  Andr\'e Nicolai$^1$, and Itzhak Roditi$^1$}
\affiliation{$^1$CBPF   Rio de Janeiro,  Brazil}
\affiliation{$^2$Department of Physics,  Cornell University, Ithaca, NY}

\bigskip\bigskip\bigskip\bigskip

\begin{abstract}
 
We study both Bose and Fermi gases at finite temperature and density  in an approximation that sums 
an infinite number of many body processes that are reducible to 2-body scatterings.   This is done for arbitrary negative
scattering length,  which interpolates between the ideal and unitary gas limits.     In the unitary limit,  
we compute the first four virial coefficients within our approximation.    The second virial coefficient is exact,
and we extend the previously known result for fermions to bosons,  and also for both bosons and fermions
for  the upper branch on the other side of unitarity (infinitely large positive scattering length).   
  Assuming bosons can exist in a meta-stable state 
before undergoing mechanical collapse,  we map out the critical temperatures for strongly coupled  Bose-Einstein condensation 
 as a function of scattering
length.          

\end{abstract}

\maketitle

\section{Introduction}

The growing amount  of increasingly accurate data from  experiments on cold atoms \cite{kett1,bloch1,chin1,nascimb1,Werner,hori1} 
poses particularly interesting challenges for theorists to develop new methods.  This is   especially
true  for   experiments where the scattering length can be tuned to vary anywhere between 
$\pm \infty$ using Feshbach resonances.    Monte-Carlo methods have been developed sufficiently  that
excellent agreement with experiments has now been achieved \cite{ ming1,Proko,bulga1,bulga2}.
Recent reviews are \cite{Castin,DrumRev}.    Nevertheless,  the development of new analytical 
methods,  though approximate,   continues to be a worthwhile pursuit because they can afford new insights into
the underlying many-body physics.   

One such method has been developed by one of us,  and is based entirely on the zero temperature S-matrix \cite{PyeTon}.   
It is reminiscent,  in fact was modeled after,  the thermodynamical Bethe ansatz equations of Yang and Yang \cite{YangYang}.  
The ingredients are the same:   the occupation numbers are parametrized  in an ideal gas  form,  but with
the one-particle energy replaced by a pseudo-energy $\vep (\kvec)$.   The latter satisfies an integral equation 
with a kernel based on the logarithm of the 2-body S-matrix at zero temperature,  and there is a simple expression for
the free energy at finite temperature and density.   Whereas for integrable  theories in 1 spatial dimension the thermodynamic
Bethe ansatz is exact because of the factorizability of the many-body S-matrix,  the formalism in \cite{PyeTon}  is certainly
an approximation.   Nevertheless it has certain desirable features,  such as the fact that the 2-body S-matrix can be
calculated exactly in non-relativistic theories,    and has been demonstrated to  give reasonably good results in some regimes.  
For instance,  it was applied to the so-called unitary limit  in 3 dimensions where the scattering length diverges,  and
the S-matrix becomes simply $-1$,  and reasonable results were obtained for the critical temperature 
 \cite{PyeTonUnitary1,PyeTonUnitary2}.   
The ratio of the viscosity to entropy density was also calculated using this method\cite{viscosity}  and agrees well
with the most recent experiments\cite{Thomas}.      
Thus,  although the method cannot really compete with numerical methods such as Monte-Carlo,  it can be justified as an exploratory tool for
regimes that have not been extensively studied.

This paper is mainly concerned with using  the method to study   the critical properties of Bose and Fermi gases 
in 3 dimensions as a function of scattering length,  including the vicinity of the unitary limit where it diverges.     
For 2 component fermions this is the familiar  BEC/BCS cross-over.   For negative scattering length the interactions
are attractive and there is a phase transition to a strongly coupled version of superconductivity.    For positive scattering 
length the fermions have a bound state,  i.e. the `atoms' form  `molecules',  which can subsequently Bose-Einstein condense.   This 
fermionic case   has been  already extensively studied
and we have nothing novel to report here.

On the other hand the bosonic case has been much less studied theoretically and is just beginning to be explored
experimentally,   and this is the main subject of this article.   The spectrum is analogous to the fermionic case:
for negative scattering length there is no bound state,  whereas for positive scattering length  molecules can form 
via 3-body processes.    We thus will restrict our study to the case of negative scattering length.     This case has perhaps not 
been studied very much theoretically because it is believed that the attractive interactions lead to a mechanical 
instability,  i.e. the gas collapses.    However it remains possible that this state could exist as a meta-stable one \cite{Ketterle}.   

 On the other side of unitarity,   in our analysis we would
need to incorporate the molecules,  with their own pseudo-energy etc,  and this is beyond the scope of this work. 
   However for  the so-called `upper branch', 
the molecules are assumed to be absent,  and this situation has been studied experimentally \cite{kett2, jochim1, strina1} 
and theoretically \cite{ohashi1,ho2,strina2,ho3}.    This motivated us to present new results on the virial expansion for
both fermions  and bosons on this upper branch.

Our results are presented as follows.    In the following section we review the S-matrix and renormalization group for
the models and  present our conventions for  the coupling and
its relation to the scattering length,  which are the usual ones.    
In section III we review our method,   describing in a precise way what we are neglecting in the approximation, 
and how in principle to calculate the corrections.     The virial expansion is studied in section IV,  
where we reproduce the known second virial coefficent for fermions on the BCS side,  but also include
new results for bosons and for the upper branch.    Here we also calculate the third and fourth virial coefficients 
in our approximation in order to compare them with more accurate calculations and experiments.   
In section V we study the extension of the Bose-Einstein condensation of the ideal gas to the full range of 
negative scattering length, thereby mapping out how $T_c$ depends on the scattering length.   
In section IV we revisit the fermionic case,  extending the results in \cite{PyeTonUnitary2} to arbitrary negative
scattering length.

\section{Conventions,  S-matrix,  Scattering length}

The bosonic model we consider  is defined by the action
for a complex scalar field $\phi$. 
\beq
\label{bosonaction}
\mathrm{S} =  \int d^3 \xvec dt \(  i \phi^\dagger  \d_t \phi - 
\frac{ |\vec{\nabla} \phi |^2}{2m}  - \frac{g}{2} (\phi^\dagger  \phi)^2 \)
\eeq
where   positive $g$ corresponds to repulsive interactions.   
By the Galilean invariance,  
the two-body   S-matrix  depends only on the difference of the incoming momentum of the two particles $\kvec, \kvec'$:  
\beq
\label{3dS}
S_{\rm matrix} (|\kvec - \kvec'|) =   \frac{8 \pi/m g_R  - i  |\kvec - \kvec'|}
{8\pi/ m g_R  + i  |\kvec - \kvec'|}
\eeq
Unitarity of the S-matrix amounts to $S^* S = 1$.  

The momentum space integrals for the higher loop corrections are divergent and an upper cut-off
$\Lambda$ must be introduced.   In the above expression, $g_R$ is the renormalized coupling:
\beq
\label{gR}
\inv{g_R} = \inv{g} + \frac{m \Lambda}{2 \pi^2}
\eeq
   Defining 
$g= \hat{g}/  \Lambda$,  where $\ghat$ is dimensionless,  and requiring $g_R$ to be independent
of $\Lambda$ gives the beta-function:
\beq
\label{betafun}
\frac{ d \ghat}{d \ell}  =  - \ghat - \frac{m}{2\pi^2} \ghat^2 
\eeq
where $\ell =  - \log \Lambda$ is the logarithm of a length scale. 
The above beta function is exact since it was calculated from the exact S-matrix.  
One  thus sees that the theory possesses a fixed point at the negative coupling
$g_* =  -2 \pi^2 / m \Lambda$ where it becomes scale invariant.

We turn now to the scattering length $\as$.   
It can be defined as \cite{LL}:
\beq
\label{landaulif}
\lim_{ | \kvec - \kvec' | \to 0}   \frac{\delta}{| \kvec - \kvec' |}  =  - \as 
\eeq
where $\delta = -i \log S$.   From the above expression for the S-matrix,  one finds
$\delta =  -2 \arctan ( m g_R  \diffk  /8 \pi )$,  which gives
\beq
\label{scatteringlength}
\as = \frac{m g_R}{4 \pi}
\eeq
One sees that scattering length diverges 
at precisely the fixed point $g=g_*$.   Heuristically,  the loss of this length scale implies universal properties of
the free energy since it can only depend on the chemical potential and temperature.   Note the  S-matrix becomes $S= -1$.     The scattering length  
$\as \to \pm \infty$, depending on from which side $g_*$ is approached.
When $g=g_*^-$, i.e. just less than $g_*$,  then $\as \to \infty$, 
whereas when $g=g_*^+$, $\as  \to -\infty$.     For reasons described above,  this paper will mainly only 
 consider negative scattering length
$-\infty <  \as  <0$.

For fermions we  consider the two-component  model defined by the action:
\begin{equation}
\mathrm{S}   = \int d^3 \mathbf{x}  dt \left( \sum_{\alpha=\uparrow ,\downarrow} i\psi_{\alpha}^{\dagger} \partial_t \psi_{\alpha} -\frac{|
\vec{\nabla}\psi_{\alpha}|^2}{2m}   - g \, \psi_{\uparrow}^\dagger\psi_{\uparrow} \psi_{\downarrow}^\dagger\psi_{\downarrow}\right)
\end{equation}
With this convention for $g$,  the S-matrix is the same as for bosons,  eq. (\ref{3dS}),  as is the beta function and scattering length
eqs. (\ref{betafun},\ref{scatteringlength}).  
For fermions,  negative scattering length corresponds to the BCS side of the BCS/BEC crossover;  we will thus only
be working on the BCS side.

For positive scattering length,  the S-matrix has a pole signifying a bound state,  or ``molecule''.    In order to have
a smooth crossover across the unitary limit,  this bound state must be incorporated into the thermodynamics,  
and this is beyond the scope of this paper.   Thus we will be primarily studying negative scattering length where there
are no molecules.     The system on the other side of unitarity where molecules are ignored is usually referred to as
the ``upper branch''.    In certain regions of density/temperature,  the upper branch can in fact be metastable,  and has 
been realized in experiments.    We will thus present a few results on the virial expansion for the upper branch for both
bosons and fermions.

\section{S-matrix based formalism for the thermodynamics}

\begin{figure}[htb] 
\begin{center}
\hspace{-15mm} 
\psfrag{f}{$\CF$}
\psfrag{fo}{$\CF_0$}
\includegraphics[width=14cm]{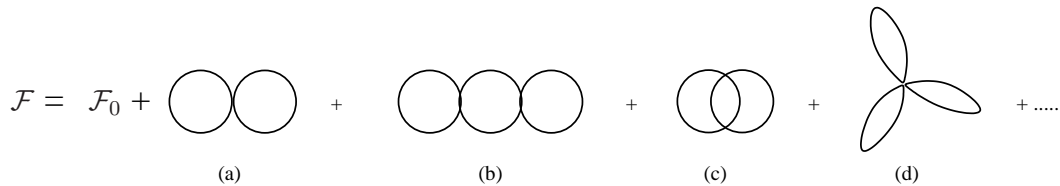} 
\end{center}
\caption{Diagrams contributing to the free energy density, where $ \CF_0$ is for the ideal gas.}  
\vspace{-2mm}
\label{Figure1} 
\end{figure}

The method developed in \cite{PyeTon} is based entirely on the S-matrix.    Contributions to the free energy
density  $\CF$  have a diagrammatic description.   Vertices with $2N$ legs represent the logarithm of the 
S-matrix for $N \to N$ particle
scattering.   Diagrams that contribute to $\CF$ are closed diagrams with vertices linked by occupation numbers
$f_0 (\kvec) = 1/ (e^{\beta (\omega_\kvec - \mu)} -s ) = z/(e^{-\beta \omega_\kvec}  - s \, z )$,  where the temperature $T=1/\beta$,  $\mu$ is the chemical potential,  $z=e^{\mu/T}$, 
$\omega_\kvec = \kvec^2 /2m $ is a 1-particle energy,  and $s=+1, -1$ corresponds to bosons, fermions respectively.  
Vertices exist for any $N$.    
These diagrams are not to be confused with finite temperature Feynman diagrams in the Matsubara formalism.
Here,  vertices represent the S-matrix to all orders in perturbation theory at zero temperature,  i.e. each vertex
already represents an infinite number of zero temperature Feynman diagrams;  the  finite temperature dependence 
comes about mainly from the occupation numbers on the legs.    This method thus appears inherently different from 
the t-matrix approach for instance.   
The low order diagrams are shown in Figure \ref{Figure1}.   Explicit expressions for some of these diagrams will
be given in the next section.   

 Two-body processes are by definition diagrams built only out of 4-vertices,  such as those in Figures 1(a,b,c).
 There are an infinite number of diagrams contributing to just the two-body processes.   An infinite subset of 
 two-body diagrams are so-called foam diagrams,  such as in Figure 1(a,b),   but with an arbitrary number of
 bubbles.    These can be resummed from a variational principal based on the diagram Figure 1a.   The result is 
 the following.     

It is convenient to express the free energy density $\CF$ and density $n$ in terms of scaling functions  $c,q$ of
the dimensionless  variables $x= \mu/T$ and $\alpha = \lambda_T / \as $,  where 
 $\lambda_T = \sqrt{2\pi/mT}$ the thermal wavelength,  as follows:
\barray
\label{freeenergy}
n \,  \lambda_T^3 &=&  q (x, \alpha)
\\ 
\CF \,  \lambda_T^3 &=&   - \zeta (5/2) T     \, c(x, \alpha) 
\earray
where $\zeta$ is Riemann's zeta function.    With the above normalizations,  for the ideal gas 
at zero chemical potential, 
$q = \zeta (3/2)$ and $c=1$.  
The two scaling functions $c$ and $q$ are of course related since
$n= - \d \CF / \d \mu$,  which leads to 
$q = \zeta (5/2) \d_x c   $.

For fermions,  is convenient to define the Fermi surface wavevector $k_F = ( 3 \pi^2 n)^{1/3}$, 
where $n$ is the 2-component density,   and $T_F =  k_F^2 /2m$.    In terms of the single component scaling function $q$:
\beq
\label{TFkF}
\frac{T}{T_F}  = \( \frac{4}{3 \sqrt\pi \, q} \)^{2/3}, ~~~~~~~\inv{k_F \as}  = \frac{\lambda_T}{\as}  ( 6 \pi^2 q )^{-1/3}
\eeq
The definitions leading to the above formulas  make sense also for  bosons;   for instance,  the 
BEC transition of the ideal Bose gas occurs at  $T_c/T_F = (4/3 \sqrt{\pi} \zeta(3/2))^{2/3} = 0.4361$.

In our formalism,  the filling fractions,  or 
occupation numbers,  are parameterized in terms of a pseudo-energy $\vep (\kvec )$   in an ideal gas form:
\beq
\label{dens}
n = \int \frac{d^3 \kvec}{(2\pi)^3} ~ \inv{ e^{ \vep  (\kvec )/T  } -1 }
\eeq
The pseudo-energy can be thought of as a 1-particle energy in the presence of all the
other (interacting) particles in the gas.    
The consistent summation of many body processes that involve only  2-body scattering  described above leads to 
an integral equation for the 
 pseudo-energy $\vep (\kvec)$,   analogous to the Yang-Yang \cite{YangYang}  integral equation.      It is convenient to define the
quantity:
\beq
\label{ydef}
y (\kvec ) = e^{- (\vep(\kvec) - \omega_\kvec + \mu )/T}
\eeq
  Then $y$ satisfies the integral
equation 
\beq
\label{inteqn}
y (\kvec ) =  1 + \inv{T}   \int  \frac{d^3 \kvec'}{(2\pi)^3} \, 
G_s (\kvec - \kvec' )  \frac{y(\kvec' )^{-1}}{e^{ \vep (\kvec') /T } -s}
\eeq
where $G_{+/-}$ refers to boson/fermion. 

The kernel $G$  is related to the logarithm of the 2-body S-matrix of the last section:
\beq
\label{Gs}
G_+ (\kvec, \kvec') = - \frac{16 \pi}{m \diffk}  \arctan \(  \frac{m g_R}{8 \pi}  \diffk  \),  
~~~~~ G_-  =  G_+/2 
\eeq
In the unitary limit $\as  \to \mp  \infty$,  the kernel takes the simple form:
\beq
\label{GUnitary}
G_+ (\kvec, \kvec' ) =  \pm \frac{ 8 \pi^2}{m | \kvec  - \kvec' |}
\eeq
where the positive sign corresponds to infinite negative scattering length.

By rotational invariance,   $y$ is a function of  $|\kvec|^2$.  
 It is convenient to define the 
dimensionless variable $k  \equiv  \sqrt{|\kvec|^2/2mT}$. 
The angular integrals in the integral equation (\ref{inteqn}) can be performed analytically (Appendix A).  
The result is the following:
\barray
\nonumber 
y(k) &=& 1 + \frac{8}{\pi}   \int_0^\infty  d k' k'    \,  \frac{z}{e^{k'^2} -  s z y (k')} 
\Biggl\{   \frac{\alpha}{2 k \sqrt{\pi}} 
\log \[  
\frac{\alpha^2/\pi  +( k  + k' )^2}
{\alpha^2/\pi  + (k - k' )^2}
\] 
\\ 
\label{ykeqn}
&~& ~~~~ ~~~~
- \( \frac{k'}{k} +1 \)  \arctan\(  \frac{\sqrt{\pi}}{\alpha} \( k + k' \) \)  
\\
\nonumber
&~& ~~~~~~~~~~~~~~
- \(  \frac{k'}{k} -1 \)  \arctan\(  \frac{\sqrt{\pi}}{\alpha} \(k - k' \) \)  
\Biggr\} 
\earray
  The scaling functions 
then have the form:
\beq
\label{nhatsc}
q   =  \frac{4}{\sqrt\pi} \int_0^\infty  dk  \,  k^2  \, 
  \frac{  y(k) z }{e^{k^2}  -  s\,  y(k) z}  
\eeq
and 
\beq
\label{cscale}
c =  \frac{4}{\sqrt{\pi} \zeta (5/2) }   
\int_0^\infty  dk  \,  k^2  \,   \(  - s \log \( 1- s  z y(k)
 e^{-k^2} \) 
- \inv{2}  \frac{ z ( y(k) - 1 ) }{e^{k^2}  - s zy(k)} \)
\eeq
The ideal, free  gas limit corresponds to $y=1$ 
where $q=  s\Li_{3/2} (s  z)$ and  $c= s \Li_{5/2} (s z)/ \zeta(5/2)$,
where $\Li$ is the polylogarithm.    The BEC critical point of the
ideal gas occurs at $\mu=0$, i.e. $q=\zeta(3/2)$.  

Since the fermion model has two components,  in equations \ref{freeenergy}, 
$q, c \to 2q, 2c$ with eqs.( \ref{nhatsc},\ref{cscale}) still valid.  In other words, 
henceforth,  $q,c$ will refer to one of the two components.

\bigskip

We will need the entropy per particle.   The expression in  \cite{PyeTonUnitary2} must be generalized for $\: \alpha  \neq 0\:$, given the extra $T$ dependence in $\mathcal{F}$ through $\alpha$.   One has 
\begin{equation}
s = -\frac{\partial\mathcal{F}(\mu,T)}{\partial T} = -\frac{\partial\mathcal{F}}{\partial x}\frac{\partial x}{\partial T} - \frac{\partial\mathcal{F}}{\partial \alpha }\frac{\partial \alpha }{\partial T} - \frac{\partial\mathcal{F}(\mu, T, x, \alpha )}{\partial T} \:,
\end{equation}
where in the last derivative only the explicit $T$ dependence is considered.
This gives
\begin{equation*}
s = \zeta\left(\frac52\right)\: \lambda_T^{-3} \left[\frac52 c - xc' - \frac12 \alpha \dot{c}\right]\:,
\end{equation*}
where $\:\dot{c} = \partial c / \partial\alpha \:$ and $\:c' = \partial c/  \partial x\:$. 
The entropy per particle then  takes the form:
\begin{equation}			\label{s_n_generalized}
s/n = \zeta\left(\frac52\right)\: \frac1q \left[\frac52 c - xc' - \frac12 \alpha \dot{c}\right]\:
\end{equation}

\section{Virial expansion}

The virial expansion is  conventionally defined as a series expansion of $\CF$,  or the density,    in powers of $z$:
\barray
\nonumber 
- \CF  \lambda_T^3 /T &=&  \sum_{n=1}^\infty   b_n z^n  \\
\label{virial.1}
n \lambda_T^3  &=&  \sum_{n=1}^\infty  n \, b_n  z^n  
\earray
where the second relation follows from $n = - \d \CF/ \d \mu$.    In the free theory,  the series expansion 
of $s \Li_{5/2} (s z )$  gives  
$b_1^{(0)}, b_2^{(0)},  b_3^{(0)},  b_4^{(0)}, ...  =   1,  \frac{s}{4 \sqrt{2}},  \inv{9 \sqrt{3}}, \frac{s}{32} , ...$.  
Expanding the occupation number in powers of $z$, 
$f_0 =z  e^{-k^2}  + z^2 \, s e^{-2k^2} + ....$,  each diagram in Figure \ref{Figure1}  can be expanded in powers of 
$z$.  Since each internal leg corresponds to an $f_0$,   a diagram with $m$ internal legs contributes 
all $b_n$ with $n\geq m$.    Thus,  the exact $b_2$ comes from Figure 1a.      The contributions to $b_3$ 
come only from  Figures 1a and 1d,  and to $b_4$ from Figures 1a, 1b, 1c,  1d,   and an additional diagram not shown
which is ``primitive'' as in Figure 1d,  but with a vertex with 8 legs and  4 loops.

In the two-body approximation captured by 
 the integral equation of the last section,   only Figures 1a  and 1b
contribute to the first 4 virial coefficients.   Let $\CF^{(a,b)}$ denote the contributions to $\CF$ from these two diagrams. 
They are given by \cite{PyeTon}
\barray
\nonumber
\CF^{(a)}  &=& \inv{2}  \int  \dk{\kvec} \dk{\kvec' }~   f_0 (\kvec)  G_s (\kvec, \kvec') f_0 (\kvec')  
\\
\label{CFab}
\CF^{(b)} &=& \frac{s}{2T}  \int   \dk{\kvec_1}  \dk{\kvec_2}   \dk{\kvec_3}  ~  f_0 (\kvec_1)  G_s (\kvec_1, \kvec_2)  f_0(\kvec_2)^2 
G_s (\kvec_2,  \kvec_3)  f_0 (\kvec_3)  
\earray

Expanding the free occupation numbers $f_0$ in powers of $z$,  the contributions to $b_{1-4}$ from diagrams 
Figure 1a, 1b,  denoted $b_{1-4}^{(a)}$ and $ b_{1-4}^{(b)}$ respectively,   are the following:
\barray
\nonumber 
b_2^{(a)}  &=&  \frac{\lambda_T^3}{2T}  \int \dk{\kvec}  \dk{\kvec'}  e^{-\omega_\kvec /T}  e^{- \omega_{\kvec' }/T}  G_s (\kvec, \kvec') 
\\
\label{bas}
b_3^{(a)}  &=&  \frac{s \lambda_T^3}{2T}  \int  \dk{\kvec}  \dk{\kvec'}  e^{-\omega_\kvec /T}  e^{- \omega_\kvec' /T}  G_s (\kvec, \kvec') 
\( e^{-\omega_\kvec /T}  + e^{-\omega_{\kvec'} /T}  \) 
\\
\nonumber
b_4^{(a)} &=& 
 \frac{\lambda_T^3}{2T}  \int \dk{\kvec}  \dk{\kvec'}  e^{-\omega_\kvec /T}  e^{- \omega_{\kvec' }/T}  G_s (\kvec, \kvec') 
 \(  e^{-2 \omega_\kvec /T}   + e^{-2 \omega_{\kvec'}  /T}  +   e^{- (\omega_\kvec + \omega_{\kvec'})/T}  \) 
 \earray  
 whereas  $b_4^{(b)}$ involves two kernels $G$: 
 \beq
 \label{b4b} 
 b_4^{(b)} =  \frac{s \lambda_T^3}{2 T^2}  
  \int   \dk{\kvec_1}  \dk{\kvec_2}   \dk{\kvec_3}~
  e^{-\omega_{\kvec_1}/T}   e^{-2 \omega_{\kvec_2} /T}  e^{-\omega_{\kvec_3}/T} ~ 
  G_s (\kvec_1 , \kvec_2)  G_s (\kvec_2,  \kvec_3) 
  \eeq

  \subsection{Infinite {\it negative}  scattering length}

  In the unitary limit,  the kernels have the simple form eq. (\ref{GUnitary}),  and the virial coefficients are pure numbers
  by the scale invariance.    The above integrals for $b_{2,3}$ are easily performed analytically by making the 
  change of variables $\kvec_1 = \kvec-\kvec', \,  \kvec_2 = \kvec +a  \kvec'$,  where $a$ is chosen to cancel the 
  cross term in the exponential;  the result then  factorizes into two gaussian integrals.   Chosing $a=n'/n$,  one  can show:
  \beq
  \label{Inn}
  \frac{\lambda_T^3}{2 T}  \int  \dk{\kvec}  \dk{\kvec'}  \,  e^{-n \omega_\kvec/T}  e^{-n' \omega_{\kvec'}/T}    
  \, \frac{4 \pi^2}{m |\kvec - \kvec'|}   =  \inv{n n' \sqrt{n+n'}}
  \eeq
  The result for $b_{2,3}  = b_{2,3}^{(0)} + b_{2,3}^{(a)}$  is then:
  \barray
  \label{b2res}
  b_2 &=& \frac{3 \sqrt{2}}{8} , ~~ b_3 = - \frac{8 \sqrt{3}}{27}  ~~~~~~~~~({\rm fermions})
  \\ 
  \nonumber 
  b_2 &=&  \frac{9 \sqrt{2}}{8} , ~~   b_3 =  \frac{19 \sqrt{3}}{27} ~~~~~~~~~~~({\rm bosons})
  \earray
  
  \def\Erf{{\rm Erf}}
  
  Finally let us turn to $b_4$ which has contributions from $b_4^{(0)}$ and Figures 1a and 1b.  
  The contribution $b_4^{(a)}$ involves only one kernel and amounts to a sum of terms involving
  the integral eq. (\ref{Inn}),  giving $b_4^{(a)} = 11/24 $ and $ 11/12 $ for fermions and bosons respectively.
  The integral for $b_4^{(b)}$ can be simplified by noting that the $\kvec_1$ and $\kvec_3$ integrals are
  identical.   After rescaling the $\kvec's$,   for fermions one obtains
  \beq
  \label{b4.1}
  b_4^{(b)} = \inv{2 \pi^{7/2}} \int  d^3 \kvec_2 \, e^{-2 \kvec_2^2} \( \int  d^3 \kvec_1 \frac{e^{-\kvec_1^2}}{|\kvec_1 - \kvec_2|}  
  \)^2 
  \eeq
  Shifting $\kvec_1 \to \kvec_1 + \kvec_2$, and performing the angular integrals one finds that the 
 $\kvec_1$  integral in parentheses is $\pi^{3/2} \Erf(k_2)/k_2$.    
 Putting this all together,  one obtains for  $b_4 = b_4^{(0)} + b_4^{(a)} + b_4^{(b)}$:
 \barray
 \label{b4.2}
 b_4 &=&  \frac{41}{96}  - 2 \sqrt{\pi} \int_0^\infty  dk  \, e^{-2 k^2}  \Erf(k)^2  = -0.0535  ~~~~~~~~({\rm fermions})
 \\
 \nonumber 
 b_4 &=&  \frac{91}{96}  + 8  \sqrt{\pi} \int_0^\infty  dk  \, e^{-2 k^2}  \Erf(k)^2  =2.870  ~~~~~~~~~~~~~({\rm bosons})
 \earray
 
It is interesting to compare with the known results for fermions.  As expected $b_2$ agrees with the exact result
calculated by other methods\cite{HoMu},
whereas $b_{3,4}$ are not since some 3 and 4 body physics has been neglected. 
   The calculation in the  work \cite{b3Theory,Leyronas} includes 3 body physics and obtains 
$b_3 = -0.29$,  in very  good agreement with experiments \cite{nascimb1},  compared to our result $b_3 = -0.5132$.
Although the sign is correct,  this does indicate the importance of the 3 body corrections.   
The coefficient $b_4$ has been extracted from experiments \cite{nascimb1} with the result $b_4 = 0.065$,  compared to 
our $-0.0535$;    here we obtain the correct trend that the absolute value of the $b$'s is decreasing, however the sign is 
incorrect.         
Note the large value of $b_4$ for bosons verses fermions.

\subsection{The upper branch:  Infinite {\it positive}  scattering length}

Let us present results for the upper branch on the other side of unitarity.   
Here,  the kernel changes sign eq. (\ref{GUnitary}).   For fermions this leads to   
\barray 
\label{BEC} 
b_2  &=&  - \frac{5 \sqrt{2}}{8} , ~~~~~~~~~b_3 =  \frac{10 \sqrt{3}}{27}, ~~~~~~  ~~~~~~~~~~~~~~~~~~~~~~({\rm fermions})
\\
\nonumber
b_4 &=&  -\frac{47}{96}   - 2 \sqrt{\pi} \int_0^\infty  dk  \, e^{-2 k^2}  \Erf(k)^2   = -0.9702 ~~~~~~~~~~~''
\earray
and for bosons:
\barray 
\label{BECbos} 
b_2  &=&  - \frac{7 \sqrt{2}}{8} , ~~~~~~~~~b_3 =  - \frac{17 \sqrt{3}}{27}, ~~~~~~~~~~~~~~~~~~~~~~~~~({\rm bosons})
\\
\nonumber
b_4 &=&  -\frac{85}{96}   +8  \sqrt{\pi} \int_0^\infty  dk  \, e^{-2 k^2}  \Erf(k)^2  = 1.034~~ ~~~~~~~~~~~~''
\earray

\section{Analysis of Bosons:  Strongly coupled BEC}

As explained in the Introduction,  in the case of a negative scattering length, we assume the bosonic gas can exist as 
a  metastable state,   
i.e. stable against mechanical collapse,  and has 
a BEC transition smoothly connected to that of the ideal Bose gas.   
 Recall that for the ideal gas,
the BEC transition occurs at $x=0$,  leading to  $q_c = n \lambda_{T_c}^3 =  \zeta(3/2)$.    
We wish to explore $q_c$ as a function of $\alpha = \lambda_T/\as$.   

\begin{figure}[h!tb]
\centering    % figura centralizada
\psfrag{x}{$\lambda_T / \as$}
\psfrag{y}{$x_c$}
\includegraphics[width=0.6\textwidth]{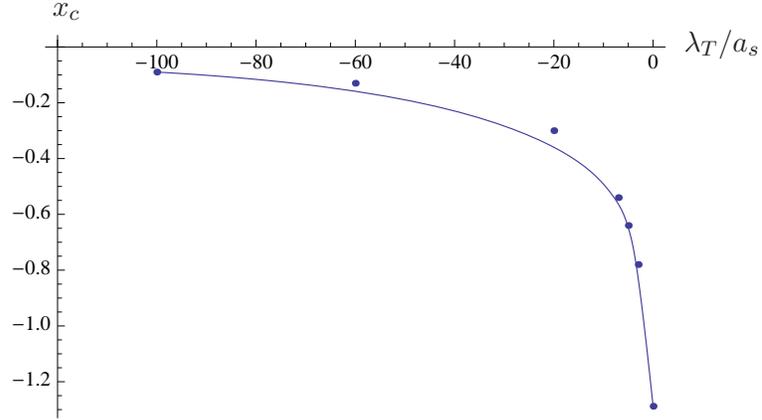}
\caption{$x_c(\alpha )$ for different values of $\alpha=\lambda_T/\as$ for bosons.  As expected,  as $\alpha \to - \infty$, 
$x_c \to 0$.   
}
\label{xcofalpha}
\end{figure}

\begin{figure}[h!tb]
\centering    % figura centralizada
\psfrag{x}{$\lambda_T / \as$}
\psfrag{y}{$ n \lambda_{T_c}^3$}
\includegraphics[width=0.7\textwidth]{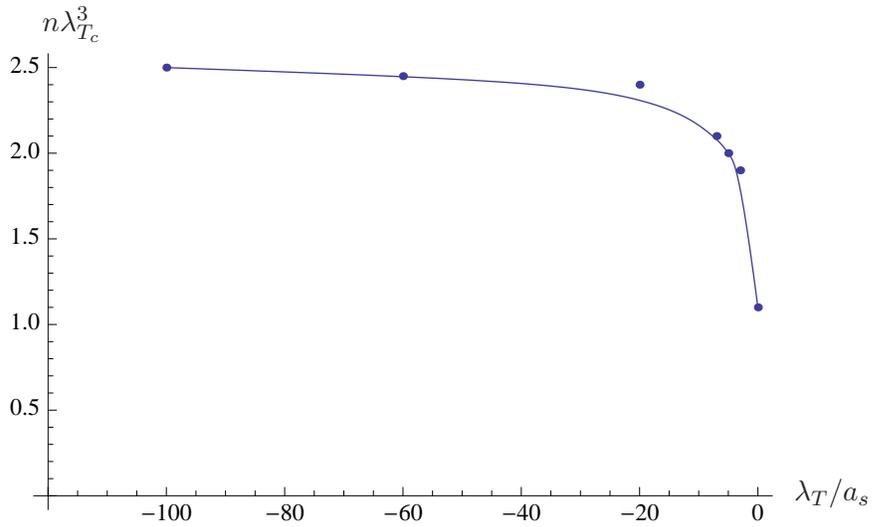}
\caption{Critical temperature (or density)  as a function of $\lambda_T/ \as $ for bosons.
}
\label{qcofalpha}
\end{figure}

\begin{figure}[h!tb]
\centering    % figura centralizada
\psfrag{x}{$1/k_F \as$}
\psfrag{y}{$T_c/T_F$}
\includegraphics[width=0.7\textwidth]{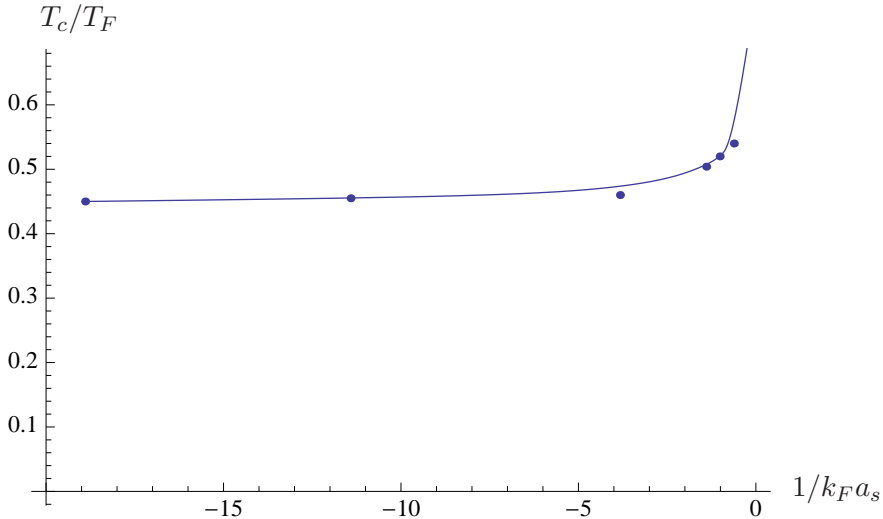}
\caption{Critical temperature in units of $T_F$ as a function of $1/k_F \as$ for bosons.
}
\label{TcTF}
\end{figure}

As for the ideal gas,  the criterion for BEC is clearly  the condition on the pseudo-energy $\vep (\kvec=0) =0$,
since this implies the occupation number $f$ diverges at $\kvec =0$.     We solved the integral equation 
(\ref{ykeqn}) iteratively until the solution converged.    A plot of $\vep(k=0)$ as a function of $x$ leads to 
the identification of the critical value $x_c (\alpha) $,  which in turn gives $q_c  (\alpha)$. 
 We plot our results for the  critical  values of $x_c $ and $q_c  = n \lambda_{T_c}^3$ in Figures \ref{xcofalpha}, \ref{qcofalpha}. 
The equivalent results in terms of $T_c/T_F$ as a function of $1/k_F \as$ are plotted in Figure
\ref{TcTF}.    In all these figures one sees that as $\as \to 0$,  one recovers the ideal gas BEC results 
$x_c=0$, $q_c = \zeta (3/2)=2.61$, and $T_c /T_F = (4/3 \zeta(3/2))\sqrt{\pi})^{2/3} = 0.436$.  
In the opposite limit $\as \to \infty$ the result is also consistent with the unitarity limit results in \cite{PyeTonUnitary2}.

\section{Analysis of Fermions}

For fermions with negative scattering length, the attractive force leads to Cooper pairing  and thus a strongly coupled
superconductivity that connects smoothly with the BCS theory at very small scattering length.    In this section we
attempt to understand this phase  transition within the present formalism.    In this case however the criterion for
the transition is not as obvious as for the BEC transition of the last section.    We will pursue the hypothesis that 
a  transition in the behavior of the entropy per particle  can be used as a signature of the phase  transition,  as was
done in \cite{PyeTonUnitary2}.

\begin{figure}[h!tb]
\centering    % figura centralizada
\includegraphics[width=0.6\textwidth]{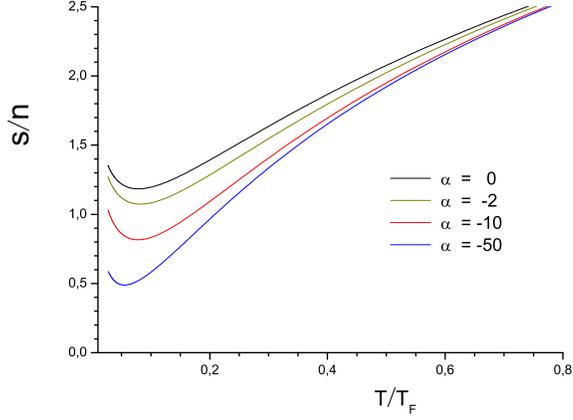}
\caption{Entropy per particle as a function of $T/T_F$.}
  \label{fig:s_n_TTF}
\end{figure}

\begin{figure}[h!tb]
\centering    % figura centralizada
\includegraphics[width=0.6\textwidth]{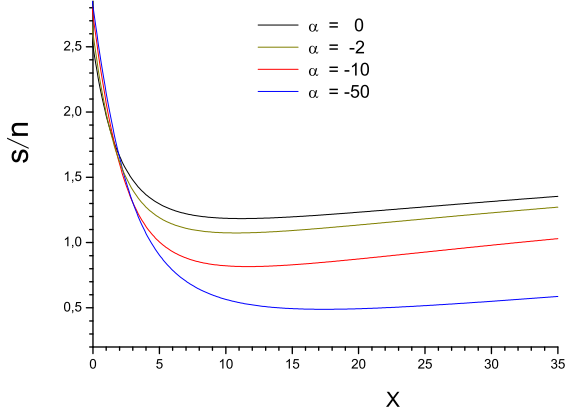}
\caption{Entropy per particle as a function of $x=\mu/T$.}
\label{fig:s_n(x)}
\end{figure}

\begin{figure}[h!tb]
\centering    % figura centralizada
\includegraphics[width=0.6\textwidth]{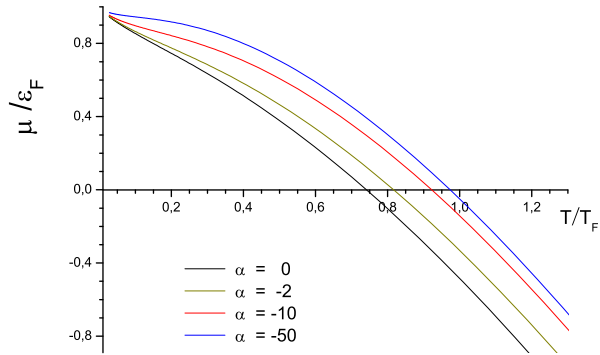}
\caption{$\mu/\epsilon_F$ as a function of $T/T_F$.}
\label{fig:mu_eF}
\end{figure}

The entropy per particle as a function of $T/T_F$ and $\mu/T$ are shown in  Figures  \ref{fig:s_n_TTF}, \ref{fig:s_n(x)},  and  Figure \ref{fig:mu_eF}  relates the chemical potential to $T/T_F$. 
  Figure \ref{fig:mu_eF}  indicates that the chemical potential $\mu$ increases with $|\alpha|$, as expected for stronger interactions.

\begin{figure}[h!tb]
\centering    % figura centralizada
\includegraphics[width=0.7\textwidth]{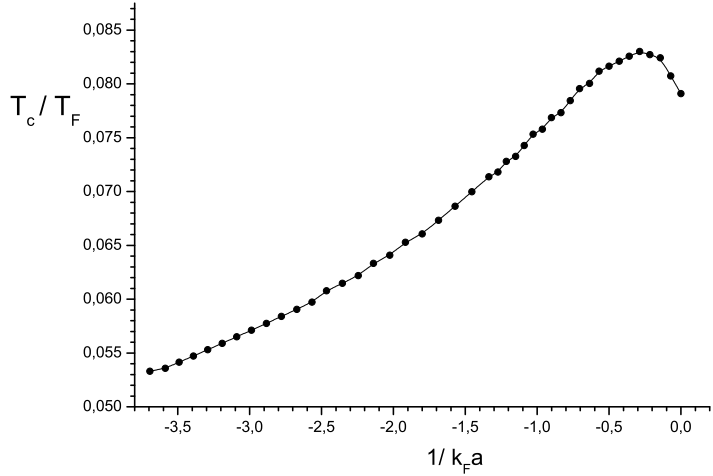}
\caption{$T_c/T_F$ as a function of $1/k_F \as$ for fermions.
}
\label{fig:crossover_S-matrix}
\end{figure}

As explained above,  we use the entropy per particle plots to identify the critical point.   One sees from Figure \ref{fig:s_n_TTF}
that the entropy begins to increase as a function of temperature when the temperature is low enough,  which is interpreted as being unphysical;   
it is actually ill defined  below this temperature, i.e. there is no longer a solution $y$ to the integral equation. 
   This change in behavior is also seen in Figure \ref{fig:s_n(x)}.
For instance,  in the unitary limit $\alpha =0^-$,  the critical point is at $x_c \approx 11.$ which corresponds to $T/T_c \approx 
0.08$.    In Figure \ref{fig:crossover_S-matrix}  we  plot our results for $T_c/T_F$ as a function of $1/k_F \as$.      Our results are in 
rough agreement with the most recent Monte-Carlo results \cite{ming1,Proko, bulga1, bulga2}.     In the range $ -0.5 < 1/k_F \as < 0$
the Monte-Carlo results are $0.08 < T_c/T_F < 0.15$,  whereas we obtain $0.07 < T_c/T_F <  0.08$,  which is on the low side.

\section{Conclusions}  

We have applied the S-matrix based formalism developed in \cite{PyeTon} to bosons and fermions 
with arbitrary negative scattering length,  extending the results in \cite{PyeTonUnitary2}  at the unitary limit.  
We explored the virial expansion up to the fourth  order within our approximation, confirming that $b_2$ for fermions 
is exactly correct, i.e. agrees with known results computed by different methods. 
   We extended these calculations to bosons and both bosons and fermions
to the other side of unitarity,  that is,  the upper branch.     Our value for $b_3$  for fermions on the other hand,  when compared with more accurate calculations and experiments,
indicates the importance of  the need to incorporate 3-body processes,  and we hope to carry this out 
in the future based on the prescription described in \cite{PyeTon}.   

 We applied the  method as an exploratory tool for less-studied regimes.   In particular,  
 for bosons,  assuming they can be  meta-stable against mechanical collapse,  we calculated
$T_c$ for a   strongly coupled Bose-Einstein condensation as a function of  (negative) scattering length.     
We hope that our results in this case  may be compared with other methods,  or better,  experiments,  in the near future.

\section{Acknowledgments}

We wish to thank  Y. Castin,  V.  Gurarie,  E.  Mueller and F.  Werner for discussions.   
   This work is supported by the National Science Foundation
under grant number  NSF-PHY-0757868 and by
FAPERJ (Funda\c{c}\~ao Carlos Chagas Filho de Amparo \`a Pesquisa do Estado
do Rio de Janeiro)
and 
CNPq (Conselho Nacional de Desenvolvimento Cient\'{\i}fico e Tecnol\'ogico).

\section{Appendix A}

For simplicity consider the bosonic case $s=+1$.  
By using
$
|\mathbf{k}-\mathbf{k}'|  = \sqrt{k^2 + (k')^2 - 2kk' \cos\theta}\;,
$
where $|\mathbf{k}| = k$, and integrating  in spherical coordinates, one has:
\begin{eqnarray}			\label{y_rho_x}
y(\mathbf{k})   &=& 1 - \frac{4 \beta }{m\pi}\int_0^{\infty} d\rho\, \rho^2 \frac{z}{e^{\beta\omega_{\mathbf{k}'}}-zy(k')} \int_0^{\pi} d\theta \frac{\sin\theta}{\sqrt{k^2 + \rho^2 - 2k\rho \cos\theta}} \nonumber  \\
			   &&  \cdotp\; \arctan \left(\frac{mg_R}{8\pi}\sqrt{k^2 + \rho^2 - 2k\rho \cos\theta}\right) \nonumber  \\
			  			  &=& 1 + \frac{4 \beta }{m\pi}\int_0^{\infty} d\rho\, \rho^2 \frac{z}{e^{\beta\omega_{\mathbf{k}'}}- zy(k')} \int_1^{-1}  \frac{dx}{\sqrt{a - b x}} \nonumber  \\
			   &&  ~~~~~\cdotp\; \arctan \left(\frac{mg_R}{8\pi}\sqrt{a - b x}\right) \: ,
\end{eqnarray}
where $\rho \equiv |\mathbf{k}'|$, $a \equiv k^2 + \rho^2\,$ , and $\:b \equiv 2k\rho\,$. In the third line above, the function $y(\mathbf{k})$ is considered to be spherically symmetric ($y(\mathbf{k}) = y(k)$). 

Evaluating the second integral
\begin{align*}	
&\int_1^{-1}  \frac{dx}{\sqrt{a - b x}} \arctan \left(\frac{mg_R}{8\pi}\sqrt{a - b x}\right) \\ 
&= \frac1{k\rho} \left[-(k+\rho)\arctan \left(\frac{mg_R}{8\pi}(k+\rho)\right) + (k-\rho)\arctan\left(\frac{mg_R}{8\pi}(k-\rho)\right)\right. \\ 
&\qquad\; \left.+\frac{4\pi}{mg_R} \log \left[\frac{1+\left(\frac{mg_R}{8\pi}\right)^2(k+\rho)^2}{1+\left(\frac{mg_R}{8\pi}\right)^2(k-\rho)^2}\right]\right]\:,
\end{align*}
one obtains:
\begin{eqnarray*}		\label{y_ro}	
y(k) &=& 1 + \frac{4 \beta }{m\pi}\int d\rho\,\rho\frac z{e^{\beta\rho ^2/2m}- zy(\rho)}\,\frac1{k} \left\{\frac{4\pi}{mg_R}\:  \log \left[\frac{\left(\frac{8\pi}{mg_R}\right)^2+(k+\rho)^2}{\left(\frac{8\pi}{mg_R}\right)^2+(k-\rho)^2}\right]\right.  \\
	  && \left. - (k+\rho)\arctan\left(\frac{mg_R}{8\pi}(k+\rho)\right) + (k-\rho)\arctan\left(\frac{mg_R}{8\pi}(k-\rho)\right) \right\}\,.
\end{eqnarray*}
Rescaling $k\to \sqrt{2mT} k,  \rho \to \sqrt{2mT} \rho$, 
 and introducing $\alpha =  \lambda_T / \as$, one finds that 
the integral equation of $y(\kappa)$ can be written as
in eq. (\ref{ykeqn}). 

The eq.  (\ref{ykeqn})   is valid for any value of $g_R$.   As a consistency check,   one can verify that in the 
unitary limit $g_R \:\rightarrow\: \pm\:\infty\,\Rightarrow\,\alpha \:\rightarrow\: 0^{\pm}\:,$ one obtains 
\begin{equation}				\label{y_unitario}
y(\kappa) 
         =  1 \mp 4 \int d\kappa' \frac z{e^{\kappa'}- zy(\kappa')} \left[\Theta(\kappa-\kappa')\sqrt{\frac{\kappa'}{\kappa}} + \Theta(\kappa'-\kappa)\right]\:   
\end{equation}
where $\kappa = k^2$, 
which is  equivalent to expression (27) of \cite{PyeTonUnitary2}.

\end{document}